# Efficient conversion of orbital Hall current to spin current for spin-orbit torque switching

Soogil Lee[1,†], Min-Gu Kang[1,†], Dongwook Go[2,3], Dohyoung Kim[1], Jun-Ho Kang[4], Taekhyeon Lee[4], Geun-Hee Lee[4], Nyun Jong Lee[5], Sanghoon Kim[5], Kab-Jin Kim[4], Kyung-Jin Lee[4], and Byong-Guk Park[1,*]

[1] *Department of Materials Science and Engineering and KI for Nanocentury, KAIST, Daejeon 34141, Korea*

[2]*Peter Grünberg Institut and Institute for Advanced Simulation, Forschungszentrum Jülich and JARA, 52425 Jülich, Germany*

[3]*Institute of Physics, Johannes Gutenberg University Mainz, 55099 Mainz, Germany*

[4]*Department of Physics, KAIST, Daejeon 34141, Korea*

[5]*Department of Physics, University of Ulsan, Ulsan 44610, Korea*

[†]These two authors equally contributed to this work.

*Correspondence to: bgpark@kaist.ac.kr (B.-G.P.)

## Abstract

**Spin Hall effect, an electric generation of spin current, allows for efficient control of magnetization. Recent theory revealed that orbital Hall effect creates orbital current, which can be much larger than spin Hall-induced spin current. However, orbital current cannot directly exert a torque on a ferromagnet, requiring a conversion process from orbital current to spin current. Here, we report two effective methods of the conversion through spin-orbit coupling engineering, which allows us to unambiguously demonstrate orbital-current-induced spin torque, or orbital Hall torque. We find that orbital Hall torque is greatly enhanced by introducing either a rare-earth ferromagnet Gd or a Pt interfacial layer with strong spin-orbit coupling in Cr/ferromagnet structures, indicating that the orbital current generated in Cr is efficiently converted into spin current in the Gd or Pt layer. Furthermore, we show that the orbital Hall torque can facilitate the reduction of switching current of perpendicular magnetization in spin-orbit-torque-based spintronic devices.**

## Introduction

Spin Hall effect (SHE) that creates a transverse spin current by a charge current in a non-magnet (NM) with strong spin-orbit coupling (SOC)[1] has received much attention because the resulting spin-orbit torque (SOT) offers efficient control of magnetization in NM/ferromagnet (FM) heterostructures of various spintronic devices[2–9]. Similar to the SHE, the orbital Hall effect (OHE) generates an orbital current, a flow of orbital angular momentum[10–13]. The OHE has distinctive features compared to the SHE; first, the OHE originates from momentum-space orbital textures, so it universally occurs in multi-orbital systems regardless of the magnitude of SOC[12]. Thus, non-trivial orbital current can be generated even in 3*d* transition metals with weak SOC[13]. Second, theoretical calculations show that orbital Hall conductivity is much larger than spin Hall conductivity in many materials, including those commonly used for SOT such as Ta and W[10,11,13]. This suggests that the spin torque caused by the OHE, or orbital Hall torque (OHT)[14,15], can be larger than the SHE-induced spin torque, enhancing the spin-torque efficiency in spintronic devices. Recently, several experimental reports have claimed that significant SOT observed in FM/Cu/oxide structures, in which the SHE is known to be negligible, is of orbital current origin[16–18]. However, there seems to be no consensus on whether those results provide evidence of OHT. One issue is that there is no exchange coupling between orbital angular momentum ($L$) and local magnetic moment, and thus the orbital current cannot directly give a torque on magnetization. To make the OHT exert on the local magnetic moment of the FM, the $L$ must be converted to the spin angular momentum ($S$)[14,15]. Therefore, finding an efficient method of "*L*-*S* conversion" is crucial to utilizing the OHT for the manipulation of the magnetization direction.

In this Article, we experimentally demonstrate two effective *L*-*S* conversion techniques of engineering SOC of either an FM or an NM/FM interface. We employ Cr as an orbital current source material because it has been theoretically predicted that Cr has a large orbital Hall conductivity $\sigma_{\mathrm{OH}}^{\mathrm{Cr}}$ of ~8,200 ($\hbar/e$)($\Omega\cdot$cm)$^{-1}$ while having a relatively small spin Hall conductivity $\sigma_{\mathrm{SH}}^{\mathrm{Cr}}$ of $-130$ ($\hbar/e$)($\Omega\cdot$cm)$^{-1}$ with the opposite sign[13]. Here, $\hbar$ is the reduced Planck constant and $e$ is the electron charge. For the Cr/FM bilayers, overall charge-to-spin conversion efficiency, referred as effective spin Hall angle $\theta_{\mathrm{SH}}^{\mathrm{eff}}$, is expressed as[14]

$$\theta_{\mathrm{SH}}^{\mathrm{eff}} = (2e/\hbar)(\sigma_{\mathrm{SH}}^{\mathrm{Cr}} + \sigma_{\mathrm{OH}}^{\mathrm{Cr}}\eta_{L-S})/\sigma_{xx}^{\mathrm{Cr}}, \quad (1)$$

where $\sigma_{xx}^{\mathrm{Cr}}$ is the electrical conductivity of Cr and $\eta_{L-S}$ is the *L*-*S* conversion coefficient.

Here, we assume perfect transmission ($T_{\mathrm{int}}$=1) of both spin and orbital currents through the Cr/FM interface. Note that the second term on the right side of Eq. (1) corresponds to the OHE contribution to $\theta_{\mathrm{SH}}^{\mathrm{eff}}$, which depends on the magnitude and sign of $\eta_{L-S}$. We demonstrate how to achieve a large $\eta_{L-S}$ by engineering SOC of an FM or an NM interfacial layer. First, we employ a rare-earth FM of Gd with strong SOC, which increases the OHT in Cr/Gd heterostructures by ten times compared to that in Cr/Co heterostructures, indicating that the orbital current generated in Cr is efficiently converted to spin current in the FM Gd layer. Second, we modify the Cr/$Co_{32}Fe_{48}B_{20}$ (CoFeB) interface by inserting a 1 nm Pt layer to facilitate *L*-*S* conversion. This leads to an enhancement in OHT, allowing us to demonstrate OHT-induced magnetization switching of perpendicular magnetization in Cr/Pt/CoFeB heterostructures. Since the OHE is expected to occur generally in various materials, our results demonstrating the significant OHT generated through the *L*-*S* conversion techniques broaden the scope of material engineering to improve spin-torque switching efficiency for the development of low-power spintronic devices.

## Results and Discussion

### Orbital Hall torque generated by orbital current in Cr

To demonstrate the OHE in Cr and associated OHT, we investigate the current-induced spin-torque in Cr/FM heterostructures for two different FMs of Co and Ni. Figures 1a,b illustrate the role of $\eta_{L-S}$ in $\theta_{\mathrm{SH}}^{\mathrm{eff}}$ of the Cr/FM samples, where $S_{\mathrm{SH}}$ is the spin angular momentum generated by SHE and $S_{\mathrm{OH}}$ is the spin angular momentum converted from orbital angular momentum due to OHE ($L_{\mathrm{OH}}$). Note that we assume that Ni has a greater $\eta_{L-S}$ than that of Co ($\eta_{L-S}^{\mathrm{Ni}} > \eta_{L-S}^{\mathrm{Co}}$) because $\sigma_{\mathrm{SH}}^{\mathrm{Ni}}$ is an order of magnitude larger than $\sigma_{\mathrm{SH}}^{\mathrm{Co}}$[19,20]. This is also supported by a recent first principle calculation demonstrating that the W/Ni bilayer exhibits a positive $\theta_{\mathrm{SH}}^{\mathrm{eff}}$ despite the negative $\sigma_{\mathrm{SH}}^{\mathrm{W}}$ of W[15]. This is attributed to the increased orbital current contribution to $\theta_{\mathrm{SH}}^{\mathrm{eff}}$ by the large and positive $\eta_{L-S}^{\mathrm{Ni}}$. Figure 1a shows the case of a Cr/Co bilayer with $\eta_{L-S}^{\mathrm{Co}}$~0, where $S_{\mathrm{SH}}$ is dominant and thus $\theta_{\mathrm{SH}}^{\mathrm{eff}}$ is mainly determined by $\sigma_{\mathrm{SH}}^{\mathrm{Cr}}$ of negative sign. On the other hand, for the Cr/Ni bilayer having sizable $\eta_{L-S}^{\mathrm{Ni}}$, non-negligible $S_{\mathrm{OH}}$ caused by the conversion of $\sigma_{\mathrm{OH}}^{\mathrm{Cr}}$ contributes to $\theta_{\mathrm{SH}}^{\mathrm{eff}}$ (Fig. 1b). Since $S_{\mathrm{OH}}$ is a positive value ($\eta_{L-S}^{\mathrm{Ni}} > 0$ & $\sigma_{\mathrm{OH}}^{\mathrm{Cr}} > 0$), opposite to $S_{\mathrm{SH}}$, $\theta_{\mathrm{SH}}^{\mathrm{eff}}$ of the Cr/Ni heterostructures becomes positive when the magnitude of $S_{\mathrm{OH}}$ is larger than that of $S_{\mathrm{SH}}$. To test whether the orbital current generated in Cr gives rise to OHT, we perform in-plane harmonic Hall measurements of Co (3.0 nm)/Cr (7.5 nm) and Ni (2.0 nm)/Cr (7.5 nm) Hall-bar patterned samples (Fig. 1c). Figures 1d,e show representative 2$^{\mathrm{nd}}$ harmonic Hall resistance ($R_{xy}^{2\omega}$) versus azimuthal angle ($\varphi$) curves under different external magnetic fields ($B_{\mathrm{ext}}$). $R_{xy}^{2\omega}(\varphi)$ is expressed as[21],

$$R_{xy}^{2\omega}(\varphi) = \{[R_{\mathrm{AHE}}^{1\omega}(B_{\mathrm{DLT}}/B_{\mathrm{eff}}) + R_{\nabla T}]\cos\varphi + [2R_{\mathrm{PHE}}^{1\omega}(B_{\mathrm{FLT}} + B_{\mathrm{Oe}})/B_{\mathrm{ext}}](\cos^3\varphi - \cos\varphi)\}, \qquad (2)$$

where $R_{\mathrm{AHE}}^{1\omega}$ and $R_{\mathrm{PHE}}^{1\omega}$ are the 1$^{\mathrm{st}}$ harmonic anomalous Hall and planar Hall resistances, respectively; $B_{\mathrm{DLT}}$ ($B_{\mathrm{FLT}}$) is the damping-like (field-like) effective field; $B_{\mathrm{eff}}$ is the effective magnetic field, including the demagnetization field and anisotropy field of FM; $R_{\nabla T}$ is the thermal contributions, and $B_{\mathrm{Oe}}$ is the current-induced Oersted field. The $R_{\mathrm{AHE}}^{1\omega}$ and $R_{\mathrm{PHE}}^{1\omega}$ data are shown in Supplementary Note 1. Figure 1f shows the $\cos\varphi$ component of $R_{xy}^{2\omega}$ divided by $R_{\mathrm{AHE}}^{1\omega}$ [$R_{\cos\varphi}^{2\omega}/R_{\mathrm{AHE}}^{1\omega}$] as a function of $1/B_{\mathrm{eff}}$ for the two FM/Cr samples, of which the slope represents $B_{\mathrm{DLT}}$ and associated $\theta_{\mathrm{SH}}^{\mathrm{eff}}$. We find that the Co/Cr sample shows a negative slope, and thus a negative $\theta_{\mathrm{SH}}^{\mathrm{eff}}$. This is consistent with the negative $\sigma_{\mathrm{SH}}^{\mathrm{Cr}}$ reported both theoretically and experimentally[22–24]. In contrast, the Ni/Cr sample exhibits a positive slope, indicating a positive $\theta_{\mathrm{SH}}^{\mathrm{eff}}$. The sign reversal of $\theta_{\mathrm{SH}}^{\mathrm{eff}}$ in the Ni/Cr sample is attributed to the increased contribution of the orbital current in Cr by the *L*-*S* conversion in Ni ($\sigma_{\mathrm{OH}}^{\mathrm{Cr}}\eta_{L-S}^{\mathrm{Ni}} > 0$). Note that the $(\cos^3\varphi - \cos\varphi)$ component of $R_{xy}^{2\omega}$ divided by $R_{\mathrm{PHE}}^{1\omega}$, representing $B_{\mathrm{FLT}} + B_{\mathrm{Oe}}$, of the Ni/Cr sample is larger than that of the Co/Cr sample (Supplementary Note 2). This might also be related to the increased orbital current in the Ni/Cr sample, which, however, should be clarified through further investigation.

To verify whether the orbital current in Cr is the main cause of the measured torque, we perform two control experiments. First, we investigate the role of FM in determining $\theta_{\mathrm{SH}}^{\mathrm{eff}}$ by measuring the $R_{\cos\varphi}^{2\omega}/R_{\mathrm{AHE}}^{1\omega}$ of the Co (3 nm)/Pt (5 nm) and Ni (2 nm)/Pt (5 nm) structures, in which Cr is replaced by Pt, which has positive $\sigma_{\mathrm{SH}}^{\mathrm{Pt}}$ and $\sigma_{\mathrm{OH}}^{\mathrm{Pt}}$ [10,11,13,25]. Figure 1f shows positive slopes and corresponding positive $\theta_{\mathrm{SH}}^{\mathrm{eff}}$'s for both the FM/Pt samples, indicating that the sign change of $\theta_{\mathrm{SH}}^{\mathrm{eff}}$ in the FM/Cr samples is not due to the FM layer itself[20]. Second, we examine the interfacial contributions[26–31] to $\theta_{\mathrm{SH}}^{\mathrm{eff}}$ by measuring the Cr thickness ($t_{\mathrm{Cr}}$) dependence of the damping-like torque efficiency, $\xi_{\mathrm{DLT}} = (2e/\hbar)(M_{\mathrm{S}}t_{\mathrm{FM}}B_{\mathrm{DLT}}/J_{\mathrm{Cr}})$, for the FM/Cr samples. Here, $M_{\mathrm{S}}$ is the saturation magnetization, $t_{\mathrm{FM}}$ is the FM thickness, and $J_{\mathrm{Cr}}$ is the current density flowing in Cr (Supplementary Note 3). If the positive $\theta_{\mathrm{SH}}^{\mathrm{eff}}$ of the Ni/Cr samples is due to the interfacial effect, $\xi_{\mathrm{DLT}}$ decreases with increasing $t_{\mathrm{Cr}}$ and eventually changes its sign to negative for thicker $t_{\mathrm{Cr}}$'s where bulk Cr with negative $\sigma_{\mathrm{SH}}^{\mathrm{Cr}}$ dominates. However, this is not the case, as shown in Fig. 1g; for both FM/Cr samples, the magnitude of $\xi_{\mathrm{DLT}}$ increases with $t_{\mathrm{Cr}}$, while maintaining its sign unchanged, which demonstrates that there is no significant interfacial contribution to $\theta_{\mathrm{SH}}^{\mathrm{eff}}$ in the FM/Cr samples. These results corroborate that the OHE in Cr primarily governs the $\theta_{SH}^{eff}$ of the FM/Cr samples, providing an excellent platform to study *L*-*S* conversion engineering.

**Efficient *L*-*S* conversion through rare-earth ferromagnet Gd**

We now present two techniques to enhance the $\eta_{L-S}$ of the Cr/FM structures. First, we introduce a rare-earth FM Gd, which is expected to have a large $\eta_{L-S}$ due to its strong SOC[32,33]. Figure 2a illustrates the *L*-*S* conversion process in Cr/Gd heterostructures, where $\eta_{L-S}^{Gd}$ is negative because of its negative spin Hall angle[34]. In this case, $S_{OH}$ due to the orbital current ($\eta_{L-S}^{Gd}\sigma_{OH}^{Cr}$<0) is in the same direction as $S_{SH}$ ($\sigma_{SH}^{Cr}$<0), so they add up constructively with each other. This would result in enhanced $\theta_{SH}^{eff}$ in the Cr/Gd heterostructure compared to the Cr/Co heterostructure.

To verify this idea, we prepare Hall-bar patterned samples of Gd (10 nm)/Cr (7.5 nm) and Co (10 nm)/Cr (7.5 nm) structures and conduct in-plane harmonic Hall measurements. Note that the measurements are performed at 10 K to avoid any side effects due to the large difference in Curie temperatures between Gd (~293 K) and Co (~1,400 K). Figures 2b and 2c show the $R_{xy}^{2\omega}(\varphi)$ data measured under different $B_{ext}$'s of the Gd/Cr and Co/Cr samples, respectively, which are well described by Eq. (2) and are represented by solid curves. The $R_{AHE}^{1\omega}$ and $R_{PHE}^{1\omega}$ data are shown in Supplementary Note 1. Figure 2d shows $R_{\cos\varphi}^{2\omega}/R_{AHE}^{1\omega}$ versus $1/B_{eff}$ for the Gd/Cr and Co/Cr samples. We find two interesting points; first, both samples exhibit negative slopes, indicating $\theta_{SH}^{eff} < 0$. Second, the Gd/Cr sample has a much larger slope or $B_{DLT}$ than that of the Co/Cr sample. The estimated $\xi_{DLT}$ of the Gd/Cr sample is −0.21±0.01, which is about ten times greater than that of the Co/Cr sample (−0.018±0.002) (Supplementary Note 3). Note that $\xi_{DLT}$ of the Gd/Cr samples increases with the $t_{Cr}$ (Supplementary Note 4), indicating that $\xi_{\mathrm{DLT}}$ originates from the bulk Cr bulk, which is the orbital current in Cr. The large enhancements of $\xi_{\mathrm{DLT}}$ or $\theta_{\mathrm{SH}}^{\mathrm{eff}}$ demonstrate that the OHT contribution can be increased by introducing FMs with large $\eta_{L-S}$.

**Magnetization switching by efficient *L-S* conversion through Pt interfacial layer**

We next demonstrate another *L-S* conversion technique that modifies the NM/FM interface by inserting a Pt layer. This method has the advantage that it can be easily incorporated into perpendicularly magnetized CoFeB/MgO structures, which is a basic component of various spintronic devices[35–37]. Figure 3a illustrates the conversion process in a Cr/Pt/CoFeB structure, where the $L_{\mathrm{OH}}$ originating from Cr is converted to $S_{\mathrm{OH}}$ in the Pt layer. Since Pt has a positive $\eta_{L-S}^{\mathrm{Pt}}$ due to positive $\sigma_{\mathrm{SH}}^{\mathrm{Pt}}$, $S_{\mathrm{OH}}$ would be positive ($\sigma_{\mathrm{OH}}^{\mathrm{Cr}}\eta_{L-S}^{\mathrm{Pt}} > 0$), while $S_{\mathrm{SH}}$ is negative ($\sigma_{\mathrm{SH}}^{\mathrm{Cr}} < 0$). Thus, the OHT due to $S_{\mathrm{OH}}$ is the opposite of the spin Hall torque due to $S_{\mathrm{SH}}$. To examine the effect of Pt insertion on OHT, we perform current-induced magnetization switching experiments as schematically illustrated in Fig. 3b. Figure 3c shows switching curves as a function of pulse current density ($J_{\mathrm{pulse}}$) for Cr (10.0 nm)/Pt (0 or 1.0 nm)/CoFeB (0.9 nm)/MgO (1.6 nm) Hall-bar patterned samples. Note that an in-plane magnetic field $B_x$ of +20 mT is applied along the current direction for deterministic switching of the perpendicular magnetization[2,4,38]. The Cr/CoFeB sample shows a counterclockwise switching curve consistent with negative $\theta_{\mathrm{SH}}^{\mathrm{eff}}$, caused primarily by the SHE in Cr. Interestingly, the switching polarity is reversed by introducing a Pt (1 nm) insertion layer. The clockwise switching curve of the Cr/Pt/CoFeB sample corresponds to positive $\theta_{\mathrm{SH}}^{\mathrm{eff}}$, which is the expected sign in the OHT scenario (Fig. 3a). Note that the sign of $B_{\mathrm{DLT}}$ and associated $\theta_{\mathrm{SH}}^{\mathrm{eff}}$ for Cr (5.0 nm)/(Pt 0 or 1 nm)/CoFeB (0.9 nm)/MgO (1.6 nm) samples, obtained from the perpendicular harmonic measurement (Supplementary Note 5), is also consistent with is the switching result.

The sign change in $\theta_{\mathrm{SH}}^{\mathrm{eff}}$ can be caused by the inserted Pt itself with positive $\sigma_{\mathrm{SH}}^{\mathrm{Pt}}$. To rule out this possibility, we investigate the $t_{\mathrm{Cr}}$ dependence of the current-induced magnetization switching for the samples, where $t_{\mathrm{Cr}}$ ranges from 2.0 nm to 12.5 nm. Figure 3d shows the switching efficiency[39,40] $\xi^{\mathrm{SW}}$ $[= (2e/\hbar)(M_{\mathrm{S}} t_{\mathrm{CoFeB}} B_{\mathrm{P}} / J_{\mathrm{SW}})]$ as a function of $t_{\mathrm{Cr}}$. Here, $t_{\mathrm{CoFeB}}$ is the CoFeB thickness, $B_{\mathrm{P}}$ is the domain wall propagation field, and $J_{\mathrm{SW}}$ is the switching current density (Supplementary Note 6). Interestingly, we find that the magnitude of $\xi^{\mathrm{SW}}$ for both samples increases with increasing $t_{\mathrm{Cr}}$, while its sign remains unchanged for all $t_{\mathrm{Cr}}$'s used in this study. Since the contribution of the spin current generated from Pt to $\theta_{\mathrm{SH}}^{\mathrm{eff}}$ in the Cr/Pt/CoFeB structures will decrease with increasing $t_{\mathrm{Cr}}$, the similar thickness dependence of $\xi^{\mathrm{SW}}$ indicates that the $\theta_{\mathrm{SH}}^{\mathrm{eff}}$ of both samples predominantly originates from the Cr layer, not from the Pt interfacial layer; the SHE and OHE in Cr are the main sources of $\theta_{\mathrm{SH}}^{\mathrm{eff}}$ for the Cr/CoFeB and Cr/Pt/CoFeB samples, respectively. These results demonstrate that the OHT can be effectively modified by interface SOC engineering and is capable of switching the perpendicular magnetization.

In conclusion, we experimentally demonstrate non-trivial OHT, spin torques originating from the orbital current in Cr, by introducing two effective ways of orbital-to-spin (*L*-*S*) conversion, which is a key ingredient of OHT generation. First, we employ a rare-earth FM of Gd having a larger *L*-*S* conversion efficiency than that of conventional 3*d* FMs. This greatly improves the SOT efficiency of the Cr/Gd bilayers compared to that of the Cr/Co bilayers. Second, we introduce a Pt interfacial layer in the Cr/CoFeB bilayers to facilitate *L*-*S* conversion. This allows the OHT to control the perpendicular magnetization in the Cr/Pt/CoFeB heterostructures. Since orbital currents can occur in various materials regardless of the SOC strength, our results provide a unique strategy based on orbital currents to develop material systems with enhanced SOT efficiency.

**Methods**

**Film preparation and Hall-bar fabrication.** Bilayers of FM (Co, Ni)/Cr, FM (Co, Ni)/Pt, Gd /Cr, and Co/Cr for harmonic measurements were deposited on Si/$SiO_2$ or Si/$Si_3N_4$ substrates using DC magnetron sputtering under a base pressure of <$2.6\times10^{-5}$ Pa, while Cr/CoFeB and Cr/Pt/CoFeB structures for switching experiments were deposited on a highly resistive Si substrate using DC and RF magnetron sputtering under a base pressure of <$4.0\times10^{-6}$ Pa. An underlayer of Ta (1 nm)/$AlO_x$ (2 nm), or Ta (1.5 nm) layers were used to obtain smooth roughness; a capping layer of Ta (2~3 nm) was used to prevent further oxidation. All metallic layers and the MgO layer were grown with a working pressure of 0.4 Pa and a power of 30 W at room temperature. The $AlO_x$ layer was formed by deposition of an Al layer and subsequent plasma oxidation with an $O_2$ pressure of 4.0 Pa and a power of 30 W for 75 s. Hall-bar-patterned devices with widths of 5, 10, or 15 μm were defined using photolithography and Ar ion-milling.

**Spin-orbit torque characterization.** In-plane harmonic measurement with AC current (frequency of 11 Hz) was performed to evaluate the spin-orbit torque of the heterostructures. Both $R_{xy}^{1\omega}$ and $R_{xy}^{2\omega}$ were recorded by two lock-in amplifiers at the same time while varying the azimuthal angle ($\varphi$) under a constant external field $B_{\mathrm{ext}}$ and a current density $J_x$ of $1\times10^{11}$ A/m$^2$.

**Current-induced magnetization switching measurements.** Magnetization switching experiments were conducted by applying a current pulse (pulse width of 30 μs) with a constant external magnetic field ($B_x$) of +20 mT. The magnetization state was checked by anomalous Hall resistance ($R_{\mathrm{AHE}}$) after applying the current pulse.

**Acknowledgements**

We acknowledge fruitful discussion with Yuriy Mokrousov, Hyun-Woo Lee, Kyoung-Whan Kim and Daegeun Jo. We also thank Byoung Kook Kim at the KAIST Analysis Center for Research Advancement (KARA) for his support on the magnetic properties measurement. This work was supported by the National Research Foundation of Korea (2015M3D1A1070465, 2020R1A2C2010309, 2020R1A2C3013302).

**Author contributions**

The study was performed under the supervision of B.-G.P. S.L. and M.-G.K. fabricated samples and conducted the in-plane harmonic measurements and spin-orbit torque switching experiments with the help of D.K., J.-H. K., T.L., G.-H.L., N.J.L., and S.K. S.L., M.-G.K., and B.-G.P. performed data analysis with the help of D.G., K.-J.K., and K.-J.L. S.L., M.-G.K., and B.-G.P. wrote the manuscript with the help of all authors.

**Competing interests**

Authors declare no competing interests.

**Data availability**

The data that support the findings of this study are available from the corresponding author upon reasonable request.

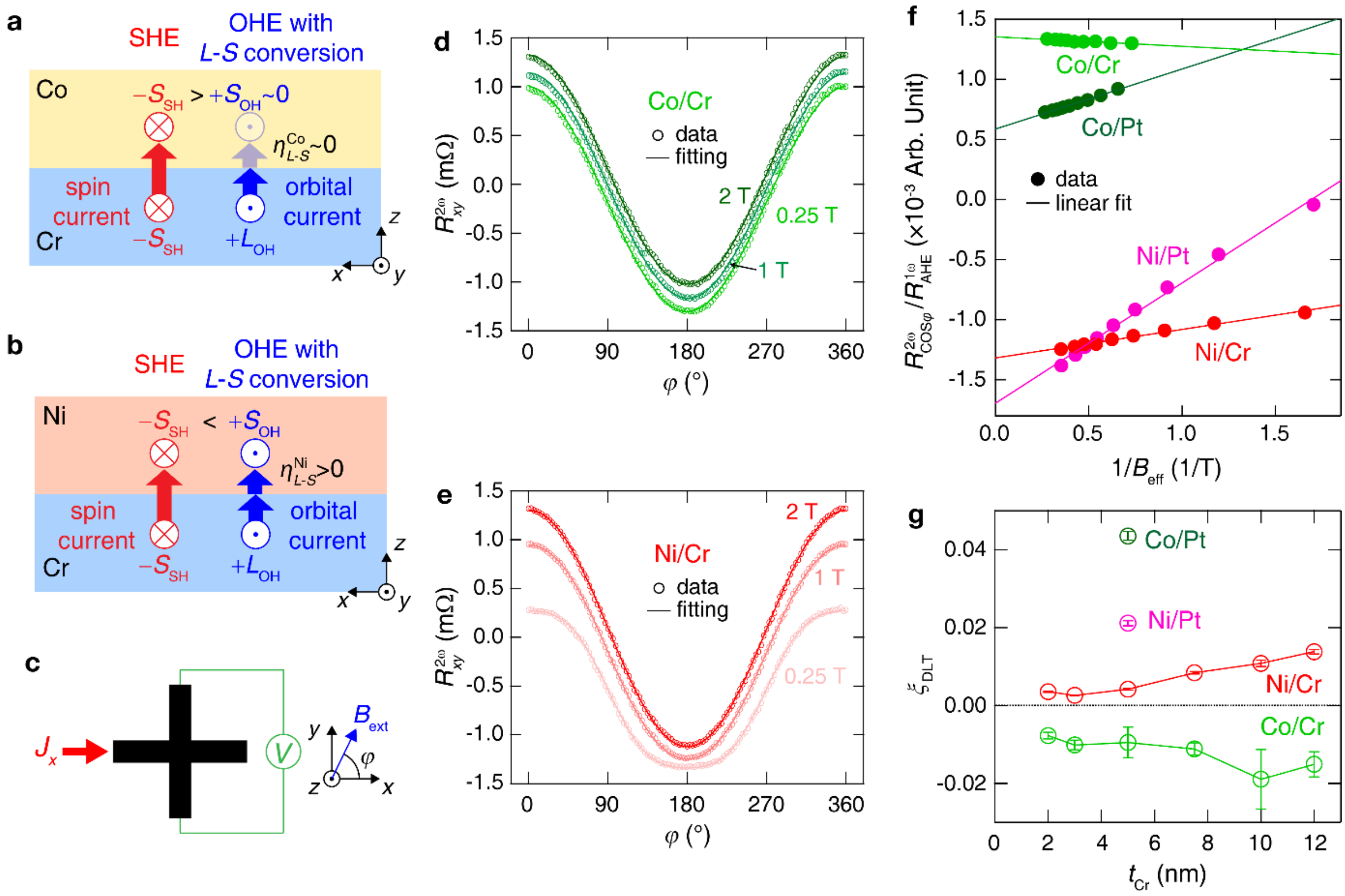

**Figure 1. Orbital-current-induced spin torque in Cr/ferromagnet heterostructures. a,b,** Schematic illustrations of the angular momentum transfer by spin current (red arrows) and orbital current (blue arrows) for Co (**a**) and Ni (**b**) FMs. The spin (orbital) angular momentum is represented by *S* (*L*). The source of *S* and *L* is marked by the subscript of SH (OH) for SHE (OHE). $\boldsymbol{\eta_{L-S}^{\mathrm{Co(Ni)}}}$ is the orbital-to-spin conversion efficiency of Co (Ni). **c,** The $\boldsymbol{R_{xy}^{2\omega}(\varphi)}$ measurement geometry for a Hall-bar sample. $\boldsymbol{\varphi}$ is the azimuthal angle of the external magnetic field ($B_{ext}$) with respect to the current direction. **d,e,** Azimuthal angle ($\varphi$) dependent 2nd harmonic Hall resistance, $\boldsymbol{R_{xy}^{2\omega}(\varphi)}$, under different $B_{ext}$ of Co (3.0 nm)/Cr (7.5 nm) (**d**) and Ni (2.0 nm)/Cr (7.5 nm) (**e**) samples. The solid lines are fitting curves using Eq. 2. Each $\boldsymbol{R_{xy}^{2\omega}(\varphi)}$ curve of Co/Cr (**d**) and Ni/Cr (**e**) is shifted by a *y*-axis offset to clearly show the $B_{ext}$ dependence. **f,** $\boldsymbol{R_{\cos\varphi}^{2\omega}/R_{\mathrm{AHE}}^{1\omega}}$ as function of $1/B_{eff}$ of Ni/Cr (red), Co/Cr (light-green), Ni/Pt (magenta), and Co/Pt (green) samples. Each solid line is a linear fitting line. **g,** Cr thickness ($t_{Cr}$) dependent damping-like torque efficiency ($\xi_{\mathrm{DLT}}$) of Ni/Cr (red) and Co/Cr (light-green) samples, where $t_{Cr}$ ranges from 2 to 12.5 nm. The $\xi_{\mathrm{DLT}}$'s of the reference Ni/Pt (magenta) and Co/Pt (green) samples are included for comparison. Lines are guide to eyes. All measurements are conducted at 300 K.

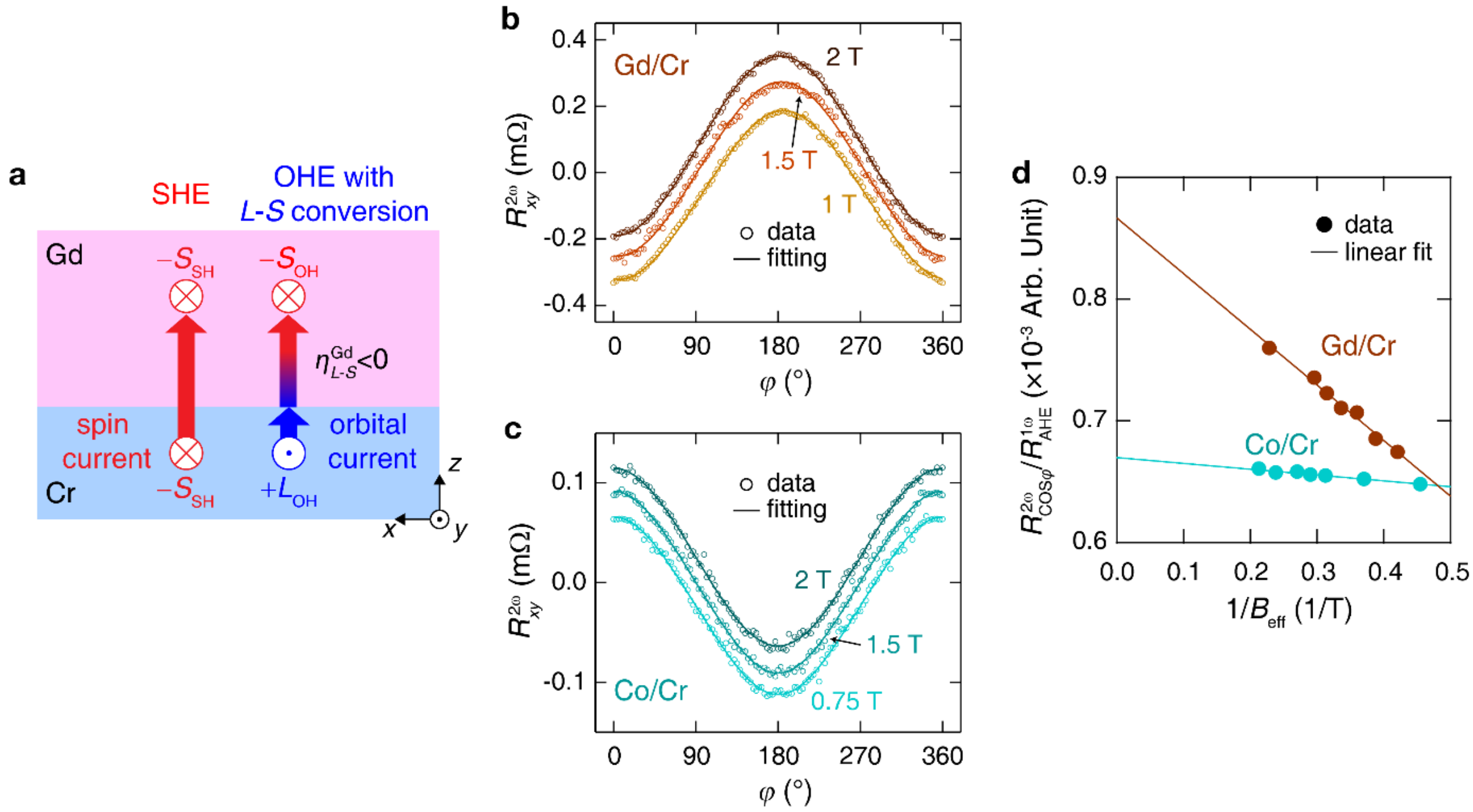

**Figure 2. Efficient *L*-*S* conversion by rare-earth Gd with strong SOC. a,** Schematic illustration of the orbital-to-spin conversion by $\eta^{\mathrm{Gd}}_{L-S}$ in the Cr/Gd heterostructure. **b,c,** $R^{2\omega}_{xy}(\varphi)$ under different $B_{\mathrm{ext}}$ of Gd (10 nm)/Cr (5 nm) (**b**) and Co (10 nm)/Cr (5 nm) (**c**) samples. Each $R^{2\omega}_{xy}(\varphi)$ of Gd/Cr (**b**) and Co/Cr (**c**) is shifted by a *y*-axis offset to clearly show $B_{\mathrm{ext}}$ dependence. The solid lines are fitting curves using Eq. (2). **d,** $R^{2\omega}_{\mathrm{COS}\varphi}/R^{1\omega}_{\mathrm{AHE}}$ vs. 1/$B_{\mathrm{eff}}$ of Gd/Cr (brown) and Co/Cr (blue-green) samples. Each solid line is a linear fitting line. All measurements are conducted at 10 K.

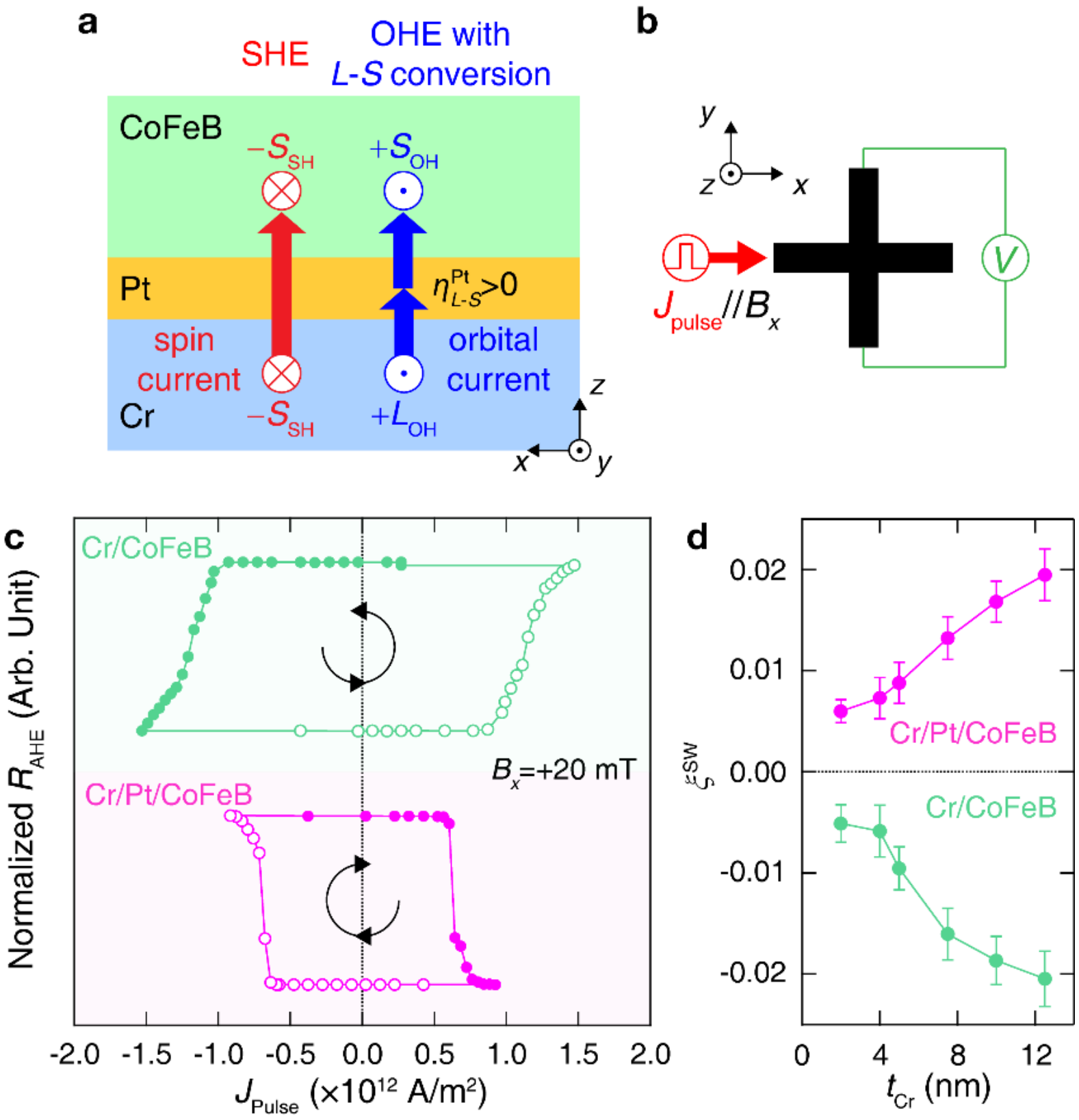

**Figure 3. OHT-driven magnetization switching by in Cr/Pt/CoFeB heterostructures. a,** Schematic illustration of the orbital-to-spin conversion by $\eta_{L-S}^{\mathrm{Pt}}$ in the Cr/Pt/CoFeB heterostructure. **b,** The magnetization switching measurement geometry for a Hall-bar sample, in which an in-plane magnetic field $B_x$ is applied along the pulsed current $J_{\mathrm{pulse}}$. **c,** Magnetization switching curves of Cr (10.0 nm)/CoFeB (0.9 nm) and Cr (10.0 nm)/Pt (1.0 nm)/CoFeB (0.9 nm) samples under a $B_x$ of +20 mT. The green and magenta symbols represent to the samples without and with the Pt insertion layer, respectively. Open (closed) symbols indicate magnetization switching from down-to-up (up-to-down) directions. The switching polarity is indicated by an arrow in the center of curve. **d,** $t_{\mathrm{Cr}}$-dependent switching efficiency ($\xi^{\mathrm{SW}}$) of the Cr/CoFeB (green) and Cr/Pt/CoFeB (magenta) samples. Lines are guide to eyes. All measurements are conducted at 300 K.